\documentstyle[12pt, graphics]{article}
\title{Field Configurations and their Instability Induced by Higher
Dimensions of Spacetime:  An Example}
\author{Laura Mersini\\Department of Physics\\University of Wisconsin-Milwaukee\\Milwaukee,
WI 53201\\lmersini@uwm.edu\\Wisc-Milw-99Th-9}
\date{21/06/99}

\begin{document}
\maketitle
\begin{abstract}
We use the model of L. Randall et al [3] to investigate the stability of
allowed quantum field configurations.
Firstly, we find that due to the topology of this 5 dimensional model, there are
2 possible configurations of the scalar field, untwisted and twisted.They give rise to two types of instability. Secondly, when allowed to interact in the brane the
untwisted field is shown to be unstable even if it is at the true
vacuum groundstate as a result of one-loop corrections that arise from
coupling with the twisted field. On the other hand, the twisted field
can make the two 3-branes (that are otherwise identical in their properties
and geometry) distinguishable therefore causing an energy difference between them. That is due to the antiperiodicity of the 
twisted fields, when rotating with $\pi$ to go from one 3-brane to the other.This energy difference between the branes renders the fifth dimension unstable.
This toy model is simple enough to use to illustrate a point that
can be important for the general case of any high dimension model, namely:
higher dimensions, besides many other effects can also induce more than
one field configuration and that can have consequences (e.g. instabilities)
even after reducing the problem to 4 dimensions.
\end{abstract}
\pagebreak
\section{Introduction.}
\setcounter{equation}{0}
The global structure of spacetime can play an important role in governing
the behaviour of quantum field theories.  We will illustrate this point
below by considering an example of a higher dimension spacetime with nontrivial
topology which reduces to a 4 dimensional flat Minkowski after imposing
boundary conditions on the 5th dimension.  The effect of the global structure
of the 5 dimensional spacetime supplemented by the boundary conditions 
will be to induce on the Minkowski submanifold more than one quantum field configuration
(in the example below it's two, in general the number of configuration will
depend on the structure of the model considered).  

We will also allow for interaction between the field configurations (besides
their respective self-interaction terms) and investigate the instability
that may arise as the result of coupling.  The idea of instabilities in
QFTh arising from nontrivial topologies was introduced by L. Ford,
D. Toms [1,2] in the context of 4 dimensional spacetime.

Our work is concerned with extending that idea to the case where nontriviality
is induced as a result of higher dimensions and is carried over even after
the reduction of the spacetime to 4 dimensions.  The model we will hire
below to illustrate this idea is the one considered by L. Randall and
R. Sundrum [3].  (In paper [3] they apply their model to a different issue,
namely mass hierarchy).  We will make use of some of their results for
the geometry of the 5 dimensional spacetime and the periodic boundary
condition of the 5th dimension $\varphi$ ($\varphi$ is an angular-like
coordinate) and see how that affects the dynamics of $\Phi^4$-quantum
fields residing on that background.

\section{The Background Metric and the Classical Solution.  A Review.}
\setcounter{equation}{0}

This section is a review of a model of a 5-dimensional spacetime of
[L. Randall, et al [3]] and their main results that we will need in
Section 3.  Consider a 4-dimensional metric multiplied by a ``warp''
factor which is a rapidly changing function of an extra dimension
$\varphi$ as follows:
\begin{equation}
ds^2 = e^{2kr_c|\varphi|} \eta_{\mu \nu} dx^\mu dx^\nu + r^2_c d\varphi^2
\end{equation}  
where $k$ is a scale of order the Planck scale, $x^\mu$ are coordinates
for the familiar 4 dimensions, while $0 \leq \varphi \leq \pi$ is the coordinate for
an angular-like extra dimension with size $r_c$.  The dynamics of particles and forces,
except gravity is confined in the 4 dimensional subspace of the above 5 dimensional spacetime,
referred to as a ``3-brane''. (They are rotationally invariant with respect
to $\varphi$.) Because our spacetime does not fill out all 5 dimensions the
boundary conditions need be specified.  These are the periodicity in
$\varphi$, the angular coordinate parametrizing the 5th dimensions, supplemented
by the identification of $(x,\varphi)$ with $(x, - \varphi)$.  The orbifold
fixed points at $\varphi = 0, \pi$ are the locations of two 3-branes, extending
in the $x^\mu$-directions so they are boundaries of the 5 dimensional spacetime.
The 3-branes can support $(3+1)$-dimensional field theories.  The 4D
components of the bulk metric are as follows:
\begin{equation}
g^{\mbox{vis}}_{\mu \nu}(x) \equiv G_{\mu\nu}(x^\mu, \varphi = \pi);\\ g^{\mbox{hid}}_{\mu \nu}(x) \equiv G_{\mu\nu}(x^\mu, \varphi = 0)
\end{equation}
where $G_{MN}$ with $M,N = \mu, \varphi$ is the $5D$ metric.  The classical
action describing the above setup is 
\begin{eqnarray}
S & = & S_{\mbox{gravity}} + S_{\mbox{vis}} + S_{\mbox{hid}}  \nonumber \\
S_{\mbox{gravity}} & = & \int d^4 x \int^\pi_\pi d \ \sqrt{-G} \left\{ - \Lambda + 2M^3 R \right\} \nonumber\\
S_{\mbox{vis}} & = & \int d^4 x \sqrt{-g_{\mbox{vis}}} \left\{ {\cal L}_{\mbox{vis}}
- V_{\mbox{vis}} \right\} \nonumber \\
S_{\mbox{hid}} & = & \int d^4 x \sqrt{-g_{\mbox{hid}}} \left\{ {\cal L}_{\mbox{hid}} -
V_{\mbox{hid}} \right\}
\end{eqnarray}
where a constant ``vacuum energy'', $\{ V_{\mbox{vis}}, V_{\mbox{hid}}\}$
is separated out of the field Lagrangians and can act as a gravitational
source even in the absence of particle excitations.  In [3] it was shown that
there exists a solution to the $5D$ Einstein's equations for the above action
with the result that 
\begin{equation}
-V_{hid} = V_{vis} = \frac{\Lambda}{k} = -24M^3k
\end{equation}
where $\Lambda$ is a cosmological constant and $M$ is the $5D$ Planck mass,
related to the $4D$ Planck mass, $M_pl$,  as follows:
\begin{equation}
M^2_{p\ell} = -\frac{M^3}{k} [ 1 - e^{2kr_c \pi} ] .
\end{equation}
Note that $g_{\mbox{hid}} = \eta_{\mu\nu}, (\varphi = 0)\; \mbox{and}\;
g_{\mbox{vis}} = e^{2kr_c \pi} \eta_{\mu\nu}, (\varphi = \pi),$ with
$\eta_{\mu \nu}$ the $4D$ Minkowski metric.  Therefore, the effective
action for a Higgs field after renormalization at the 3 brane of the field,
$\psi \rightarrow e^{k r_c \pi} \psi_0$ becomes:
\[ S_{\mbox{vis}} \supset \int d^4 x \sqrt{-g_{\mbox{vis}}} \{ g^{\mu\nu}_{\mbox{vis}}
\partial_\mu \psi^+_0 \partial_\nu \psi_0 - \lambda(|\psi_0|^2 - v^2_0)^2 \} \]
\begin{equation}
= \int d^4 x \sqrt{- \eta} e^{4 k r_c \pi} \{ \eta^{\mu\nu} 
e^{2k r_c \pi} \partial_\mu \psi^+_0 \partial_\nu \psi_0 - \lambda\left[|\psi_0|^2
- v^2_0 \right]^2 .
\end{equation}
i.e.
\begin{equation}
S_{\mbox{eff}} \supset \int d^4 x \sqrt{-\eta} \left\{ \eta^{\mu\nu}\partial_\mu
\psi^+ \partial_\nu \psi - \lambda \left[|\psi|^2 - e^{2kr_c \pi} v^2_0 \right]^2 \right\}
\end{equation} 
Thus the physical mass scales are set by the symmetry-breaking scale:
\begin{equation}
v \equiv e^{k r_c \pi} v_0
\end{equation}
\section{Field Configurations and their Instability in the 3-branes.}
\setcounter{equation}{0}
In paper [3] the authors focused their attention in the mass hierarchy
problem (Eq. 2.8).  However, the $S^1 \times R^4$, $5D$ spacetime that
they considered (Eq. 2.1) has other interesting properties.  The
property we will focus our attention on here, is the topology of scalar
fields that are allowed in that spacetime (2.1).

Consider the Lagrangian of the scalar fields to be similar to the Higgs one
in (2.6, 2.7) but also containing a linear term of the form 
$- \frac{\epsilon}{2v}(\psi + v)$, i.e. containing a false and a true vacuum
with energy difference $\epsilon$.  For example $\epsilon$ can be taken to be a part
of $V_{\mbox{vis}} = \frac{\Lambda}{k}$ as that was the constant ``vacuum energy'' 
separated out in 2.3.  Note that because of the global topology $S^1 \times R^4$
of the spacetime we can have two field configurations on each 3-brane
i.e. $\psi(x,\varphi) = \pm \psi(x,-\varphi)|_{\varphi = 0,\pi}$.  Since
the dynamics of the field is confined to the 4 dimensional subspace (i.e.
$\psi(x^\mu, \varphi)|_{\varphi=0,\pi} = e^{k r_c|\varphi|} \cdot \Psi(x^\mu)|_{\varphi=0,\pi}$)
after imposing the boundary conditions at $\varphi = 0,\pi$ we have the
following:
\[ \psi_{\mbox{hid}}(x,0) = \pm \Psi_{\mbox{hid}}(x,\varphi=-0) \]
\[ \psi_{\mbox{vis}}(x,\pi) = \pm e^{k r_c \pi}\Psi(x,\varphi= - \pi) \]
For simplicity, let's drop the indices $\{\mbox{vis,\, hid}\}$ below and
concentrate on the dynamics of real scalar fields in the visible sector (everything would apply equally to the
hidden sector). 

Let's denote the untwisted field by $\psi$, i.e.
\begin{equation}
\psi(x,\pi) = \Psi(x, - \pi)e^{kr_c \pi}
\end{equation}
on the 3-brane, and the twisted
field configuration by $\widetilde{\psi}$, i.e. 
\begin{equation}
\widetilde{\psi}(x,\pi) = - \widetilde{\Psi}(x, - \pi)e^{kr_c \pi}
\end{equation}
The renormalizable Lagrangian for those 2 fields may also involve an 
interaction  between the 2 configurations as follows:
\begin{eqnarray}
{\cal L} & = & \frac{1}{2} \left[ \partial_\alpha \psi \partial^\alpha \psi +
\partial_\alpha \widetilde{\psi} \partial^\alpha \widetilde{\psi} \right] -
\frac{\lambda_1}{4!} \left[ |\psi|^2 - v^2_1 \right]^2 + \frac{\epsilon}{2v_1}
(\psi+ v_1)\nonumber\\
& & - \frac{\lambda_2}{4!}[|\widetilde{\psi}|^2 - v^2_2]^2 + \frac{\epsilon}{2v_2}
[ - \widetilde{\psi} + v_2] - \frac{1}{2} g \psi^2 \widetilde{\psi}^2.
\end{eqnarray}  
Note from (3.1) and (3.2) that the renormalization of the two fields, $\psi$ and $\widetilde{\psi}$ involves the same expontetial factor $e^{k r_c \pi}$ resulting from the reduction of the  $\varphi$ angular coordinate at $\varphi = \pi$ In addition, the twisted field 
picks up a negative sign, i.e. in $5D$ it has the topology of a M\"{o}bius
Strip.  This sign does not affect the quartic or quadratic terms in the
Lagrangian (3.3) but it does affect the linear term , for example if both fields
are in the true vacuum, after rotation by $\pi$ in the fifth dimension $\varphi$ that takes us
to the other brane, the untwisted field finds itself again in true vacuum,
while the twisted field ends up in the false vacuum.  This is due to the fact that the linear term (the one containing $\epsilon$, the energy difference between the two vacuums)  in the lagrangian (3.3) is not invariant under the diffeomorphism $\widetilde{\psi}-> {\widetilde{-\psi}}$.(For simplicity we ignored the coupling
term  in drawing Fig.1 below)\\
\vspace*{2.0in}\\
Fig. 1:  Self interaction  potential for a) untwisted field b) twisted field.\\
\\
In the lagrangian (3.3) $v_{1,2}$ are the renormalized vacuum  values of the fields, set by the symmetry-breaking scale as given in Eq. (2.8). Thus, even if the constant ``vacuum energy'' of the untwisted configuration
$V_{\mbox{vis}}$ ,in both branes, is canceled out by the 5 dimensional cosmological
constant $\Lambda$ (as is often the case in higher dimensional theories) the vacuum energy may not be canceled  in the nontrivial case of twisted field configuration. For example ,if the twisted field in one brane happens to be in the true vacuum, (the energy of which may be zero due to cancellation by $\Lambda$), then, in the other brane,  this antiperiodic  field will  be found in the false vacuum which obviously will have nonzero energy.     
This effect due to the antiperiodicity
of the twisted fields would render the two 3-branes distinguishable. Further, the energy difference between the two 3-branes renders the fifth dimension unstable as in this case the branes will attract (collapse of the fifth dimension) or push each other away (inflating fifth dimension) depending on the sign of the energy difference.\\ In order to calculate the effect of interaction between these two fields we make use
of Ford's results in [1] obtained through zeta function methods.  Let us assume
that the coupling constants $\lambda_1, \lambda_2, g$, are small.  Also
assume $\epsilon / {v^4_{1,2}} \ll \lambda_{1,2}$ so that we can  drop
the term proportional to $\epsilon$ in the equations of motions, ( in the context of tunneling this condition on $\epsilon$ is known as the thin wall approximation).We can use perturbation methods only around the stable solutions, i.e. around the vacuum states, which from the lagrangian (3.3) are: $v_1$ for the $\psi$ field and $v_2$ for the $\widetilde{\psi}$ field. (These are the locations of true vacuum, one could equally apply the following expansion to the false vacuum states) We will assume
coherent ground states for both  fields in their respective vacuums $v_1, v_2$. Denote their  shifted values relative to the vacuum by: $\xi$= $\psi - v_1$, $\widetilde{\xi}$=$ \widetilde{\psi} - v_2$. After writing the lagrangian (3.3) in terms of the new fields $\xi$ and $\widetilde{\xi}$ we can find  perturbative solutions for these shifted fields $\xi$ and $\widetilde{\xi}$.  Then, to first order
in $\lambda_1$, $\lambda_2$ and $g$ (see Ford [1]) we have:
\begin{eqnarray}
\left\{ \begin{array}{lll} 
\partial_\mu \partial^\mu \xi + g < \widetilde{\xi}^2_0 >_0 \xi + \lambda_1
< \xi^2_0>_0 \xi  + m^2_1 \xi    & = & 0 \\
\partial_\mu \partial^\mu \widetilde{\xi} + g < \xi^2_0>_0 \widetilde{\xi} +
m^2_2 \widetilde{\xi} + \lambda_2 < \widetilde{\xi}^2_0 >_0 \widetilde{\xi} 
& = & 0 
\end{array} 
\right. 
\end{eqnarray} 
where
$\xi_0, (\widetilde{\xi}_0)$ is a free field operator which satisfies:
\begin{eqnarray}
\left\{ \begin{array}{lll}
\partial_\mu \partial^\mu \xi_0 + m^2_1 \xi_0 = 0 \\
\partial_\mu \partial^\mu \widetilde{\xi}_0 + m^2_2 \widetilde{\xi}_0 = 0 \\
\ m^2_1 = \frac{\lambda_1}{3}v^2_1 + v^2_2 g \\
\ m^2_2 = gv^2_1 + \frac{\lambda_2}{3}v^2_2 \\
\end{array}
\right.
\end{eqnarray}
Eq. (3.5) is the zeroth order expansion in the coupling constants $\lambda_{1,2}$ and $g$ and Eq.(3.4) is of the first order in that expansion.To arrive at Eq. (3.4) above, we also made use of  the relation
 $<z|\xi^3|z>$ = $<z|\xi|z>^3 +3 <\xi^2_0>_0 <z|\xi|z> + O(\lambda)$,where the coherent state $|z>$ is an eigenstate of the anihilation operator associated with the field $\xi_0$, and $<z|\xi|z>$ is required to be sufficiently small such that 
the $<z|\xi|z>^3$ may be neglected (see (12-14 in paper [1] for more details on the above relation). The symbol $< >_0$ denotes the vacuum expectation values calculated with respect to the coherent ground state $z=0$.For brevity, we used the same symbol as the field $\xi$ in Eq.(3.4) to denote the expectation value of the field with respect to $|z>$ i.e. $<z|\xi|z>$. The expectation values 
$< \widetilde{\xi}^2_0 >_0$ and $< \xi^2_0 >_0$ are calculated by
Ford [1] for the  $S^1 \times R^3$ case. They can readily be extended
for $S^1 \times R^4$ as follows:
\begin{equation}
<0|\xi^2_0|0> = \frac{e^{2kr_c\pi}}{2(2\pi)^3 L} \sum^\infty_{n = 0}
\int^\infty_0 d^3K  \cdot \omega^{-1}=\frac{e^{2kr_c\pi}}{2(2\pi)^3L}[\pi^{3/2} \frac{2\pi\Gamma(-1/2)}{L \Gamma(1)} F(-1/2;\zeta,0)]
\end{equation}   
where
\begin{equation}
K^2\; = k^2_x + k^2_y + k^2_z, \quad k_{\varphi} = \frac{2n\pi}{L} = 
\frac{n}{r_c}, \quad F(\lambda;\zeta,b)=\sum^\infty_{j=-\infty}f(j), \quad f(j)=[(j+b)^2 + \zeta^2]^{-\lambda}\;
\end{equation}
The periodicity length $L = 2 \pi r_c$, and
\begin{equation}
\omega_{1,2} = [p^2 + m^2_{1,2}]^{1/2} \quad \mbox{with} \quad
p^2 = K^2 + k^2_\varphi, \;.
\end{equation}
The function $F(-1/2;\zeta,0)$ in the above integration (3.6) can be expressed in terms of another function $f_{1/2}(\zeta)$ defined by:\\
$f_{\lambda}(a,0) =\int^\infty_a du \frac{u^2 - a^2}{e^{2\pi u} - 1}$ , (see the Appendices in [1] for a complete treatment of these functions). After substracting  from (3.6) the part that does not depend on $L$, the regularized part yields:
\begin{equation}
<\xi^2_0>_0\; = \frac{e^{2kr_c\pi}}{4 r^2_c \pi^2} f_{-\frac{1}{2}}(\zeta) \quad
\mbox{where} \quad
\zeta = \frac{m_{1,2}r_c}{2}.
\end{equation}
Similarly, one finds that  $<\widetilde{\xi}^2_0>_0 = <\xi^2_0>_{L=4r_c\pi} - <\xi^2_0>_{L=2r_c\pi}$.
The lowest expectation value of $<\xi^2_0>_0$, $<\widetilde\xi^2_0>_0$ occur when  $m_{1,2} = 0$, ($f_{-\frac{1}{2}}(0) = \frac{1}{24}$) with 
\begin{equation}
<\xi^2_0>_{m_1=0}\; =
 \frac{e^{2kr_c\pi}}{96 r^2_c \pi^2}, \quad <\widetilde{\xi}^2_0>_{m_2 = 0}\;
= \frac{- e^{2kr_c\pi}}{192 r^2_c \pi^2}
\end{equation}
The exponential warp factor that comes from the metric (view it as a conformal factor) in (3.10) shifts the expectation values and can give divergences when
 ${r_c}-> {\infty}$ in (3.10). In order to avoid this divergence, we define the exponent in the warp factor of the metric,  $kr_c\pi$ to be constant. If  $k$ is taken to be of the order of the Planck scale as in Sec.2,it then follows that $r_c$, the size of the the 5th dimension, is proportional to the  Planck size.  Of course, instead  of the above approach, one could absorb the metric exponential factor in the scaling of  the momentum $K$ in 4D as well,(in the same manner as the mass scaling of the lagrangian) i.e. treat $K e^{2k r_c \pi}$ as the physical momentum . However, the parameter $r_c$ is not neccessarily constrained to be small, and in the most general case it can take any value, including  the  limit ${r_c}-> {\infty}$.
 Thus, it makes physical sense to demand that the  exponent be a constant, in order to have finite renormalized parameters of the theory (e.g. mass, coupling constants momentum, etc.). Then, one can still have the mass hierarchy scaling as well as finite expectation values in Eq.(39-3.10), in the whole range of $r_c$, even for 
 $r_c ->{\infty}$, as long as $k ->0$.\\
The square of the frequency of the  untwisted field $\xi$ in Eq. (3.4) can become negative for certain values of the parameters, especially at  $p=0$. The result of this calculation is that the twisted field is always stable
against interactions with itself or the untwisted field ( to be precise, for as long as the coupling constants are small enough to satisfy: $\frac{e^{2kr_c\pi}(g-2\lambda)}{192\pi^2} < 1$), while from (3.4) the excitations of the untwisted
configuration can be unstable if: 
\begin{equation}
m^2_1 + g < \widetilde{\xi}^2_0>_0 + \lambda_2 < \xi^2_0 >_0 < 0 .
\end{equation}
In particular consider for simplicity $m_1 = 0$; then
\begin{equation}
\omega^2_k = [p^2 - \frac{e^{2kr_c\pi}}{192 r^2_c \pi^2} (g - 2 \lambda_2)]
\end{equation}
where $p^2 = k^2_x + k^2_y + k^2_z + k^2_\varphi$. We could equally have  written (3.12) in terms of the physical momentum shifted by the scale factor $e^{kr_c\pi}$ instead of $p^2$,  to show that after factoring out the exponent, the frequency  $\omega_p$ explicitly rescales in the same manner as the mass .
The smaller the size of the 5th dimension $r_c$ becomes, the more $p$-modes 
become unstable.  At $k_x = k_y = k_z = 0$ the untwisted field can become
unstable, as $k_\varphi = \frac{n}{r_c}$ can also be zero for $n = 0$.
While the twisted field, $k_\varphi = \frac{(2n+1)}{r_c}$,is greater
than zero even for $n=0$ i.e. there is a minimum eigenfrequency (and positive energy gap )even for
the zero mode $n=0$.

The above shows that the untwisted scalar field in the brane can be made unstable through its interaction
with the twisted field $(\omega^2_p < 0)$ even when it resides in the 
ground state of the true vacuum, i.e. its excitation can grow exponentially
via small perturbations in $\lambda_{1,2}$ and $g$.  However,if in our treatment of the twisted field we had chosen the false vacuum as its ground state then the twisted  field would not have been stable either,in fact not even at the tree level $\widetilde\xi_0$, as the result of quantum
tunneling, i.e. it would have decayed to its other vacuum ( true vacuum).  It would be interesting to see how the dynamics of a decaying twisted field (when it tunnels to another vacuum ), would affect the stability of the untwisted field. The coupling constant of their interaction $g$ multiplied by the expectation value $\widetilde{\xi_0}^2$ that enters Eq. (3.4) for the field $\xi$, would  play the role of a time-dependent mass  since the decaying twisted field in this case is a dynamic field.A thorough and consistent treatment of this problem would require that we take into
account the dynamics of the decaying (via tunneling) twisted field resulting  when solving for its interaction with the untwisted
configuration in (3.4).  This could be the subject of future investigation.
\section{Conclusions}
The formalism of [3] focused our attention on the
problem of scalar field configurations and their stability on the 3-branes
which are the boundary of a periodic $5D$ spacetime, $S^1 \times R^4$, at
$\varphi = 0, \pi$.  This spacetime is interesting because,
besides the property of mass scaling effect [3], we showed  that
the global topology of this spacetime can induce 2 field configurations
on the 3-branes, owing to the periodicity in the extra dimension
$\varphi$ with periodicity length $L = 2r_c \pi$.  
The $(3+1)$-dimensional bulk metric of the 3-branes at the boundary
$\varphi = 0,\pi$ was taken to be Minkowskian for simplicity.

The potential considered was the usual Higgs or false vacuum decay potential
of $\lambda \psi^4$-theory (Eq. 3.3) in the Minkowski spacetime which
is a submanifold of the $5D$, $S^1 \times R^4$, manifold as in equation (2.1).
As a result, the two scalar field configurations allowed were the twisted and
untwisted fields.  

Whenever the Lagrangian of the matter fields contained a linear term (3.3)
(or an odd power of the field in general) then the two 3-branes could be
made distinguishable as the result of: a)the antiperiodicity of the twisted
fields; b)the non-invariance of the linear terms (odd powers)  
under the diffeomorphism $\widetilde{\psi} \rightarrow - \widetilde{\psi}$.
In other words, there is a broken symmetry between the two 3-branes because the
twisted field has odd parity (antisymmetric) which does not commute with the  rotation operator. This energy difference between the two 3-branes, in the case of the twisted field configuration, renders the fifth dimension unstable because the condition for the energy cancellation that ensures a static fifth dimension can not be satisfied.

It was further shown that there is another kind of instability in the Minkowskian submanifold  in the general case where coupling between
the two fields is allowed,in which case the untwisted fields can be driven to instability
(especially in the low energy modes) as a result of its interaction with
the twisted fields, ( through the one-loop quantum corrections, see
Eq. 3.10-3.11.) Note that the smaller the compactification radius
$r_c$ , the more modes become unstable, (see Eq. 3.11) (One could define the inverse of the compactification radius $r_c$ in (3.12) to play the role of an  UV cutoff frequency for all the unstable modes of the untwisted field, thus considering the unstable modes as unphysical).\\ On the
other hand, the twisted fields were perfectly safe with respect to
quantum correction, but they could decay via quantum tunneling,
([4]).  There are two possibilities in trying to deal with
this instability of  untwisted fields: either the effective potential
of the untwisted field contains another local minimum that the field
can decay to; or this model prefers only twisted fields, which are  stable  and excludes the existence of untwisted fields.

Finally, we used this toy model to illustrate the idea that higher
dimensions can induce (after compactification) more than one field
configuration on the 4 dimensional subspace with significant effects. The vacuum energy for some of the configurations (e.g. twisted field) may not be cancelled out by the $5D$ cosmological constant $\Lambda$, in at least one of the branes, resulting after dimensional reduction in a $4D$ effective cosmological constant as well as distinguishability between the branes therefore instability of the model. Furthermore, one should
be aware that the mutual interaction among these different configurations in the submanifold can render some
of the fields unstable due to quantum loop corrections, even when they are taken stable at the tree level. Perhaps some criterion can be used to exclude the existence of one of the field configurations otherwise allowed by the topology of this spacetime. In that case stability of this model would be possible through careful fine-tuning of the parameters and energies of the fields.\\
\noindent\\
Acknowledgment:
I thank Prof.L.Parker  for his  continuous help and support in our weekly discussions, and for suggesting the idea of distinguishability between the branes . I want to thank Dr.A.Raval also, for our many helpful discussions.This work wassupported in part by the NSF Grant No. Phy-9507740  
\pagebreak\\
{\large{\bf References.}}\\
\begin{itemize}
\item[{1)}] L.H. Ford, Phys.Rev. D21, 933 (1980) ; Phys.Rev. D22, 3003 (1980)
\item[{2)}] D.J. Toms, Phys.Rev. D21, 928 (1980); Phys.Rev. D21, 2805 (1980) 
\item[{3)}] L. Randall and R.Sundrum, 'A large mass hierarchy from a small extra dimension', $(hep-ph/9905221)$ 
\item[{4)}] See Phys.Rev. D59, 123521 (1999), ($hep-th/9902127$) and $references$ therein for the treatment of tunneling and false vacuum decay.
\end{itemize}

\resizebox{3in}{3in}{\includegraphics{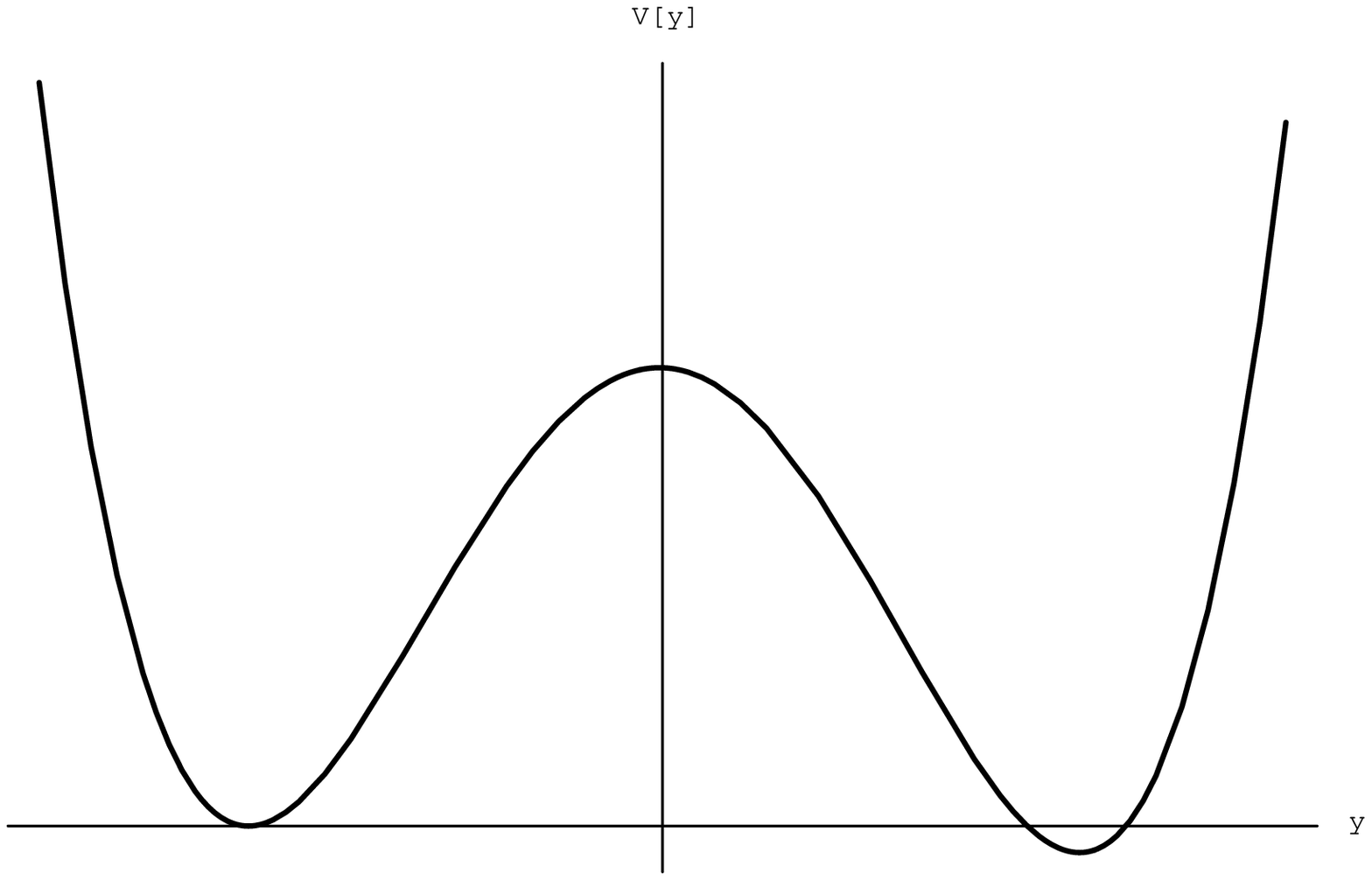}}
\begin{center}
1.a
\end{center}

\resizebox{3in}{3in}{\includegraphics{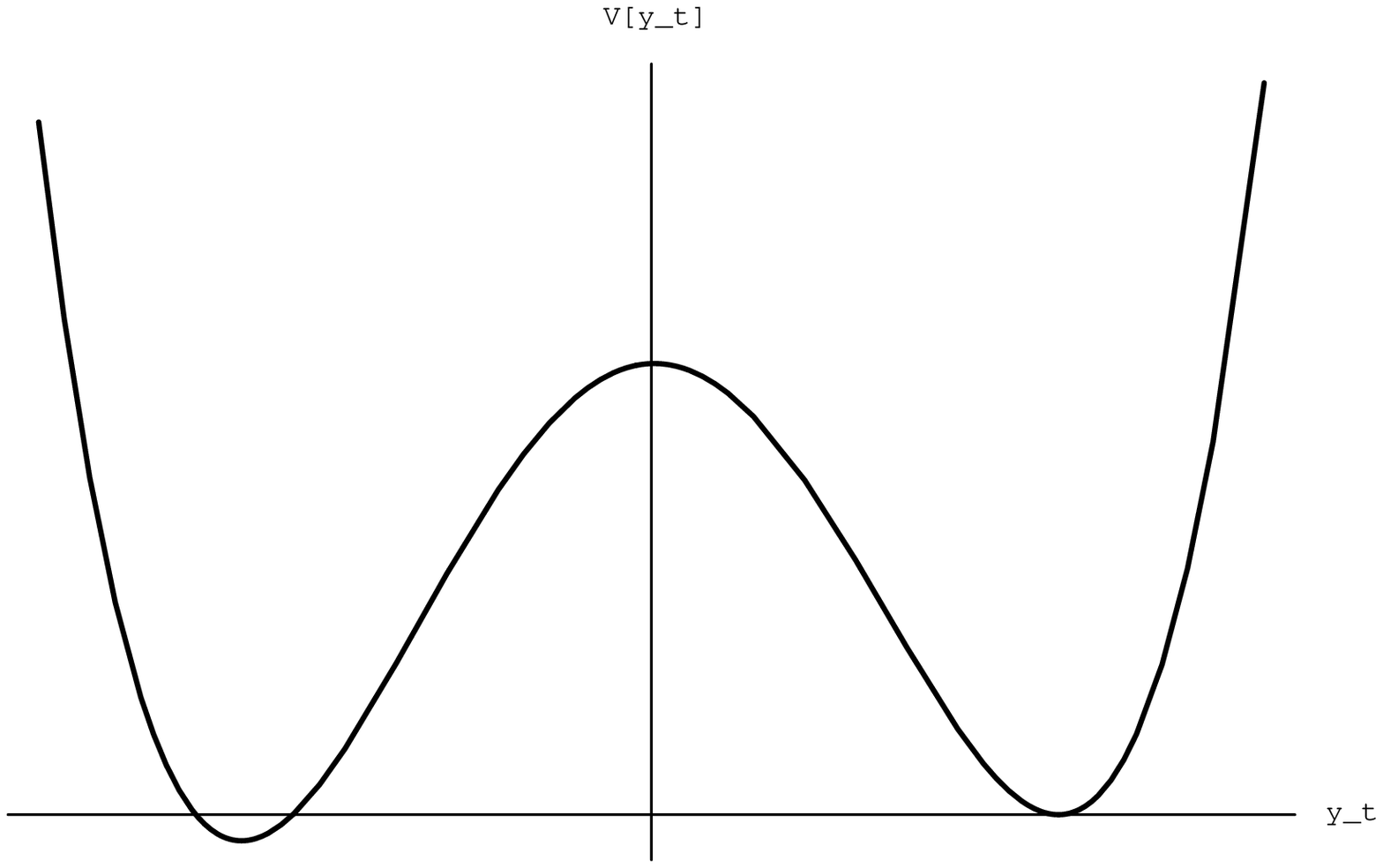}}
\begin{center}
1.b
\end{center}

\begin{center}
Fig. 1 Self interaction potential for a) the untwisted field, b) twisted field
\end{center}

\end{document}